\documentclass[12pt,letterpaper]{article}
\usepackage[utf8]{inputenc}
\usepackage[T1]{fontenc}
\usepackage{lmodern,textcomp}
\usepackage[letterpaper,margin=1in]{geometry}
\usepackage{microtype,etoolbox}
\usepackage{amsmath,amssymb,amsthm,mathtools,bm}
\usepackage{xcolor,tikz}
\usetikzlibrary{arrows.meta,positioning,calc}
\usepackage{booktabs,array,adjustbox,longtable,float,pdflscape,graphicx}
\graphicspath{{figures/}}
\usepackage{caption}
\usepackage{setspace}
\usepackage[authoryear,round]{natbib}
\let\cite\citet
\usepackage[hidelinks]{hyperref}
\hypersetup{
  pdftitle={Weak-Instrument Inference and Model Diagnostics in Double Machine Learning for Linear Instrumental Variable Models},
  pdfauthor={Airui Meng},
  pdfkeywords={Double machine learning, Partially linear model, Weak instruments, Identification-robust inference, Model selection}
}
\makeatletter
\renewcommand{\@seccntformat}[1]{%
  \csname the#1\endcsname\ifstrequal{#1}{section}{.}{}\hskip 0.5em}
\makeatother

\input{supplementary_numbers.tex}
\providecommand{\tightlist}{\setlength{\itemsep}{0pt}\setlength{\parskip}{0pt}}
\theoremstyle{plain}
\newtheorem{theorem}{Theorem}
\newtheorem{proposition}{Proposition}

\theoremstyle{definition}
\newtheorem*{diagnostic}{Diagnostic characterization}
\title{\LARGE Weak-Instrument Inference and Model Diagnostics in Double Machine Learning for Linear Instrumental Variable Models}
\author{Airui Meng\thanks{Teachers College, Columbia University,
525 West 120th Street, New York, NY 10027, USA.
Email: \href{mailto:am6817@tc.columbia.edu}{am6817@tc.columbia.edu}.}}
\date{October 2026}
\begin{document}
\maketitle
\begin{abstract}
\normalsize\singlespacing\noindent
We study weak-instrument-robust inference in partially linear
instrumental-variables models with learned nuisances. Building on an
i.i.d.\ uniform-size baseline, we establish uniform asymptotic validity
of clustered cross-fitted Anderson--Rubin inference and cross-fitted
score inference under the stated sampling and nuisance-rate conditions.
Extensions establish Anderson--Rubin validity for survey-weighted target
laws and fold-nested generated instruments. A supporting population
benchmark shows that prediction-optimal instrument residualization
maximizes concentration when the treatment nuisance is correctly
specified. A five-design diagnostic portfolio, a known-truth Monte
Carlo, and an exploratory selector examine how nuisance specification
and estimation affect realized strength and inference.
\end{abstract}

\bigskip
\section{Introduction}\label{introduction}

Double machine learning (DML) enables estimation and inference in
partially linear instrumental-variables (PLIV) models with flexible,
high-dimensional controls. Under strong identification and suitable
regularity and nuisance-estimation rate conditions, cross-fitted
orthogonal scores deliver asymptotically normal estimators and valid
conventional inference \citep{chernozhukov2018}. When residualized
identification is weak, these guarantees need not apply. The central
problem is to obtain valid confidence sets after nuisance estimation
and selection without requiring a strong residualized first stage.

Applied sampling designs introduce additional requirements. With
clustered observations, score variation and nuisance-estimation errors
must be controlled at the cluster level. Survey weights can change
the target law and the conditional means used for residualization.
A generated instrument adds construction error and must be estimated
without using its inference fold. Validity for independent observations
and an observed instrument under an unweighted target law does not by
itself establish validity in these settings.

Learner sensitivity motivates a related diagnostic question. In
CollegeDistance, \cite{cetin2026a} finds that regularized linear learners
produce estimates close to conventional two-stage least squares,
whereas random forests produce near-zero estimates with small Wald
standard errors. \cite{cetin2026a} introduces residualized concentration
under strong identification and proposes absorption of identifying
variation as an explanation. Two questions guide our analysis: how to
conduct inference when residualized identification is weak, and how to
interpret strength differences across nuisance learners.

We address these questions in two steps. The main inferential results
establish uniform asymptotic validity of clustered cross-fitted
Anderson--Rubin inference and cross-fitted score inference under the
stated conditions. AR extensions cover survey-weighted target laws
and fold-nested generated instruments. A supporting population
benchmark holds the instrument and conditioning set fixed and takes
the treatment nuisance at its true conditional mean. In that setting,
prediction-optimal instrument residualization maximizes concentration,
providing a reference for interpreting realized learner differences.

For inference, we use the Anderson--Rubin test \citep{andersonRubin1949} and related identification-robust procedures \citep{kleibergen2005,moreira2003}, which control size under weak identification. We also assess realized first-stage strength using a robust Wald statistic, which coincides with the \cite{montielOleaPflueger2013} effective \(F\) in the single-instrument case.\footnote{See
also \cite{lee2022} and \cite{andrewsStockSun2019}.}
Applying these tests to cross-fitted PLIV moments requires control of
errors from nuisance fitting and selection, together with a consistent
covariance estimator under the relevant sampling design. These
requirements underlie the validity results below; the effective-\(F\)
statistic serves as a diagnostic of realized strength.

We make three contributions, centered on the validity of
identification-robust inference after nuisance learning.

\begin{itemize}
\tightlist

\item
  \textbf{Weak-instrument-robust inference for applied PLIV designs.}
  We derive the local-to-zero limit of the just-identified cross-fitted
  estimator and its feasible Wald statistic, clarifying why a small
  standard error does not establish reliable precision under weak
  identification (Proposition~\ref{prop:wald}).
  Within the leakage-free nested cross-fitting framework of
  \cite{cetin2026b}, we establish uniform asymptotic validity of
  the clustered Anderson--Rubin procedure and the cross-fitted score
  statistic under the stated sampling and nuisance-rate conditions
  (Theorems~\ref{thm:clustered-ar} and~\ref{thm:score}).
  We further develop extensions for survey-weighted target laws
  and fold-nested generated instruments (Propositions~\ref*{prop:weighted}--\ref*{prop:generated}).
  The generated-instrument extension covers AR inference; its score
  counterpart is established in the single-instrument case.
  Monte Carlo experiments evaluate finite-sample coverage and the
  informativeness of the resulting confidence sets, including
  stress settings with coverage shortfalls.

\item
  \textbf{Population optimality of instrument residualization.}
  We establish a population benchmark for interpreting sensitivity to
  nuisance learners. Holding the instrument and conditioning set fixed,
  and taking the treatment nuisance at its true conditional mean,
  the prediction-optimal instrument residualization maximizes
  residualized concentration over all admissible residualizations
  (Proposition~\ref{prop:concentration}). Recovering the instrument's conditional mean
  therefore entails no population concentration loss. Together with
  estimand invariance under the maintained homogeneous-coefficient
  PLIV model (Proposition~\ref{prop:invariance}), this result distinguishes changes in
  identification strength from changes in the parameter being estimated
  and provides a benchmark against which observed learner differences
  can be assessed.

\item
  \textbf{Empirical diagnosis of learner sensitivity.}
  Across four of the five applications, we combine nuisance-fit diagnostics,
  residualized first-stage strength, and identification-robust
  confidence sets to investigate why estimates and inference vary
  across learners.
  The evidence supports distinct diagnoses: overfit dilution in the
  examiner design, geographic fold leakage and coupled nuisance
  misspecification in CollegeDistance, and reduced conditional
  relevance along a fixed-sample, fixed-instrument Card control
  ladder.
  A randomized application provides a stable-strength benchmark.
  Known-truth simulations isolate the role of treatment-nuisance
  specification: the relative strength ranking of flexible and
  regularized linear residualizations reverses between correctly
  specified and misspecified treatment nuisances.
  These comparisons show how apparently similar empirical patterns
  can call for different changes to nuisance estimation,
  cross-fitting, or the interpretation of the conditional design.

\end{itemize}

Together, these results show why learner choice, residualized instrument strength, and inference must be
considered jointly in DML-IV applications.

\subsection{Related literature}\label{related-literature-and-contribution}

The analysis builds on \cite{cetin2026a,cetin2026b}.
\cite{cetin2026a} introduces residualized concentration and documents
learner sensitivity under strong identification; \cite{cetin2026b}
provides leakage-free nested cross-fitting (LF-NCF), identification-aware
cross-validation (IACV), and i.i.d.\ uniform-size AR theory.\footnote{The nested selection idea, product-rate architecture,
concentration measure, CollegeDistance sensitivity and proposed
mechanism, restricted-\(\pi\) analysis, and Angrist--Krueger stress
exercise originate in \cite{cetin2026a}. \cite{cetin2026b} also supplies
the tuning-overfit discussion and the ridge-AR and conditional-QLR
results drawn on here.}
Our inferential contribution concerns the additional validity arguments
needed for cluster dependence, survey-weighted target laws, and
fold-nested instrument construction.\footnote{The i.i.d.\ AR result of \cite[Theorem~3]{cetin2026b}
is restated as Theorem~\ref*{thm:iid-import} in Supplementary \ref*{a.4-proof-of-theorem-1}.}
\cite{cetin2026c} studies power when a scalar AR direction is learned
from a growing instrument dictionary. Here the PLIV moment is fixed
and its nuisance functions are selected; our theory establishes size
and coverage for that moment under the sampling designs considered.

\cite{ma2026} is the closest published study. It develops an
orthogonalized Anderson--Rubin test for the LATE with
high-dimensional covariates and extends the analysis to scores
that are linear in the parameter, including the PLIV score,
through a cross-fitted conditional QLR test
(Appendix B, Example 1). The two approaches share a cross-fitted,
Neyman-orthogonal construction and an emphasis on uniform
asymptotic validity. Ma's results are established for i.i.d.\
observations, an observed instrument, an unweighted target law,
and cross-fitted nuisance learners satisfying its regularity and rate
conditions. These conditions allow flexible learners; the study does
not separately develop selection guarantees for a learner library.
Our analysis uses the nested selection framework described above and
establishes validity for cluster dependence, survey-weighted target
laws, and fold-nested instrument construction. These settings require,
respectively, control of cluster-sum remainders, orthogonality under
the weighted law, and bounds on instrument-construction error.

An early antecedent of selection-aware IV inference is
\cite{belloni2012}, who use lasso
to select optimal instruments from a high-dimensional dictionary
and provide a sup-score test that is robust to weak identification.
The distinction lies in the object of selection: their procedure
selects instruments, whereas our analysis concerns the nuisance
learners used to residualize the outcome, treatment, and instruments.
\footnote{The sup-score procedure of \cite{belloni2012}
uses raw instruments after low-dimensional linear partialling,
provides pointwise size control, and does not cross-fit the
three PLIV nuisance functions.}

A complementary literature develops inference that is robust to
the number of instruments. Building on \cite{chaoSwanson2005}
and \cite{hansenHausmanNewey2008}, \cite{mikushevaSun2022}
construct a jackknife Anderson--Rubin test when the instrument
count grows proportionally with sample size. \cite{navjeevan2025}
develops a dimension-agnostic jackknife score statistic, and
\cite{limWangZhang2024} propose a dimension-agnostic bootstrap
Anderson--Rubin procedure. These studies treat the instruments
as fixed, remove controls by exact linear projection, and assume
independent observations. \cite[p.~3, fn.~3]{navjeevan2025} identifies
accommodating high-dimensional controls while incorporating
first-stage information as an open issue. \cite{zhangLiSun2026} combine cross-fitted orthogonal moments and multi-task learning and propose a continuous updating estimator and corresponding Wald statistics that are robust to increasing IV dimensions and high-dimensional covariates. Our orthogonalized,
cross-fitted PLIV construction addresses this issue in a
complementary setting: the instrument dimension is fixed and
moderate, while machine learners accommodate high-dimensional
or nonparametric control functions.
\footnote{The ``cross-fit'' variance estimators of \cite{mikushevaSun2022} are leave-one-out variance proxies within a
full-sample projection statistic. They differ from the fold-based
sample splitting used here to accommodate machine learning.
Supplementary \ref*{app:d6} discusses their pre-test interpretation
of the quarter-of-birth design.}

For clustered data, \cite{wangZhang2024} establish the validity
of wild bootstrap Anderson--Rubin inference in linear IV models
with few large clusters. Their asymptotic framework complements
the regime of many moderate clusters in Theorem~\ref{thm:clustered-ar}. 

Recent panel working papers examine related questions in
longitudinal settings.
\cite{baiardi2026} use first differences and apply a panel analogue of the first-stage \(F\) statistic and an Anderson--Rubin test to shift-share migration designs.
\cite{avilaMarquez2026} shows that linear reduced-form
specifications can make panel instruments spuriously weak
and proposes a Super Learner control-function estimator
with a generalized concentration parameter.
\footnote{Both studies remain unpublished working papers
as of this writing.}

The examiner-design literature provides another setting in which
instrument construction and nuisance learning interact.
These designs use the leniency of a quasi-randomly assigned
decision-maker as an instrument for treatment
\citep{dobbie2018,chyn2025}. Our high-dimensional application uses a fold-nested cluster-jackknife leniency instrument and examines how flexible residualization affects its realized strength.

Finally, a parallel literature on flexible IV methods studies
different target parameters. \cite{athey2019}
and \cite{hartford2017} target heterogeneous effects or
structural functions, while \cite{blandhol2026} study the implicit weighting of the linear IV
estimand under flexible controls. These objects differ from
the scalar homogeneous-coefficient PLIV parameter maintained
here. A like-for-like comparison would therefore first require
testing the homogeneity restriction.

The remainder of the paper is organized as follows.
Section~\ref{model-and-identification} presents the PLIV model and identification results,
and Section~\ref{estimation-and-inference} develops estimation and inference.
Section~\ref{monte-carlo-studies} evaluates coverage and nuisance specification
in Monte Carlo studies. Section~\ref{empirical-applications} presents the applications
through a common diagnostic protocol, comparative evidence, and a
cross-application synthesis. Section~\ref{discussion} discusses the implications and
limitations and concludes. The Supplementary Material contains proofs,
full result tables, and simulation designs (Supplementary \ref*{appendix-a-formal-results-sources-and-proofs}--\ref*{app:c}),
together with application, selection, and cross-fitting details
(Supplementary \ref*{app:d}--\ref*{app:f}).

\section{Model and Identification}\label{model-and-identification}

\subsection{Setup}\label{setup}

We study the partially linear instrumental-variables (PLIV) model \[
Y = \tau_0\,D + g_0(W) + \zeta, \qquad \mathbb{E}[\zeta \mid Z, W] = 0,
\] where \(D\) is a scalar endogenous regressor, \(Z\) is an instrument
of dimension \(d_z\), \(W\) is a vector of controls, and \(\tau_0\) is
the scalar structural parameter of interest. Estimation rests on the
residualized, Neyman-orthogonal PLIV score \[
\psi(\tau, \eta) = \big(Z - \pi(W)\big)\big[\,(Y - \mu(W)) - \tau\,(D - r(W))\,\big],
\] where \(\eta = (\mu, r, \pi)\) collects the nuisances, whose true
values (subscript \(0\)) are \(\eta_0 = (\mu_0, r_0, \pi_0)\) with
\(\mu_0(W) = \mathbb{E}[Y \mid W]\),
\(r_0(W) = \mathbb{E}[D \mid W]\), and
\(\pi_0(W) = \mathbb{E}[Z \mid W]\). We write residualized variables as
\(\tilde Y = Y - \mu_0(W)\), \(\tilde D = D - r_0(W)\), and
\(\tilde Z = Z - \pi_0(W)\). A naming caution: the classical ``first
stage'' regresses the treatment on the instruments, and \(\pi_0\) is
not that object. We call \(\pi_0\) the
\textbf{instrument-residualization nuisance} (the residualization, for
short). The classical first stage appears here only in residualized
form, as the \(\tilde D\)-on-\(\tilde Z\) relation whose strength §\ref{concentration-governs-precision-and-the-failure-mode-is-distributional}
and §\ref{design-a-structured-diagnostic-portfolio-in-residualized-strength} measure.

We take as given the leakage-safe nested cross-fitting of \cite{cetin2026a,cetin2026b}.
Nesting matters because, under weak identification, selection that
reuses inference observations can distort the null law of
identification-robust statistics at first order. The
protocol splits the complement of each held-out (inference) fold into
an estimation and a selection subsample. Candidate learners for
\((\mu, r, \pi)\) are fit on the estimation subsample, selected on the
selection subsample, and the selected fits evaluate the held-out fold,
with no refit on the union. Every selected object is therefore
measurable with respect to data disjoint from the fold it scores. Under a product-rate condition and strong identification, \cite[Theorem~2]{cetin2026a} delivers \(\sqrt n\)-asymptotic normality of the
residualized-IV estimator. Our
contribution begins where the strong-identification assumption is relaxed. Before examining inference under weak identification, we first clarify whether the choice of instrument residualization changes the population estimand.

\subsection{Estimand invariance across instrument residualizations}
\label{the-estimand-does-not-depend-on-the-choice-of-consistent-first-stage}

Different instrument-residualization learners can produce different
estimates, but this does not by itself establish that they target
different population parameters. To interpret this sensitivity,
we first clarify the role of \(\pi\) in the population PLIV moment.
With the outcome and treatment nuisances at their true conditional
means, the maintained exogeneity condition ensures that any
admissible instrument residualization leaves the moment valid
at \(\tau_0\). The following proposition establishes the
corresponding estimand invariance and consistency result.

\begin{proposition}[Estimand invariance and the role of the
instrument residualization]\label{prop:invariance}

Work at a fixed population under the sampling and moment conditions
in Supplementary \ref*{a.1-proposition-1-sources-and-proof-sketch}.
Suppose the PLIV model holds with
\(\mathbb E[\zeta\mid Z,W]=0\) and
\(g:=\mathbb E[\tilde Z\tilde D]\ne0\).
For any square-integrable, \(W\)-measurable residualization limit
\(\pi_\ast\), assume
\(\Omega_\ast:=\mathbb E[(Z-\pi_\ast(W))(Z-\pi_\ast(W))']\succ0\).
If the cross-fitted fits \((\hat\mu,\hat r,\hat\pi)\) converge to
\((\mu_0,r_0,\pi_\ast)\) in the held-out empirical \(L^2\) norm
defined there, the residualized-IV estimator is consistent for
\(\tau_0\). The limit \(\pi_\ast\) need not equal \(\pi_0\).
Under these conditions, the choice of instrument residualization
can affect residualized concentration without changing the
estimand or consistency. 

\end{proposition}

One implication is that a linear instrument residualization can
yield a consistent estimator even when \(\pi_0\) is nonlinear:
the linear projection remains an admissible limit \(\pi_\ast\).
Thus, under the stated conditions, changing the
instrument-residualization learner does not change the probability
limit of the estimator. Differences between finite-sample estimates
may nevertheless reflect estimation error or differences in
residualized strength. If the outcome or treatment nuisance is
misspecified, the conditions supporting this invariance no longer
hold. A near-zero estimate under flexible learning therefore
does not, by itself, establish either a zero structural effect
or a loss of identification.

These conclusions concern the maintained homogeneous-coefficient
PLIV model. Under effect heterogeneity, the estimated coefficient
may instead represent a weighted average of heterogeneous responses,
whose causal interpretation requires separate justification.
The target map in Section~\ref{design-a-structured-diagnostic-portfolio-in-residualized-strength} records the statistical
and causal targets maintained in each application.

Proposition~\ref{prop:invariance} concerns consistency at a fixed population with
nonzero residualized relevance. Supplementary \ref*{a.1-proposition-1-sources-and-proof-sketch} provides the proof and states the scope of the associated precision comparisons. We next examine how weak
residualized identification affects the estimator's sampling
distribution and the validity of conventional Wald inference.

\subsection{Wald inference under weak residualized identification}\label{concentration-governs-precision-and-the-failure-mode-is-distributional}

Define the per-observation residualized concentration \[
C = g_{ZD}'\,\Omega^{-1} g_{ZD}, \qquad g_{ZD} = \mathbb{E}[\tilde Z \tilde D], \quad \Omega = \mathbb{E}[\tilde Z \tilde Z'],
\] and the concentration parameter \(\mu^2 = nC\). The classical
concentration parameter divides \(nC\) by the first-stage error
variance and is dimensionless; our \(\mu^2\) keeps the treatment
scale and enters only order and ranking statements within one design. The asymptotic
variance of the residualized-IV estimator scales as
\(\mathrm{Var}(\hat\tau) \propto 1/(nC) = 1/\mu^2\) under conditional
homoskedasticity of the score; under bounded conditional
heteroskedasticity \(C\) ranks weak-identification strength up to
constants (Supplementary \ref*{a.1-proposition-1-sources-and-proof-sketch}), so a
reduction in \(C\) is a loss of identification strength at fixed \(n\).

The inverse-concentration variance approximation relies on
strong identification. When residualized relevance is weak,
the validity of the usual normal approximation requires
separate examination. In particular, a near-zero estimate
with a small reported Wald standard error need not establish
a precisely estimated zero. Proposition~\ref{prop:wald} formalizes this
distinction in the just-identified case. It first establishes
that rescaling the residualized instrument leaves the Wald
variance unchanged, and then characterizes the estimator
and feasible Wald statistic under a local-to-zero sequence.
The key issue is whether the statistic has the reference
distribution required for conventional Wald inference.

\begin{proposition}[Scale invariance and Wald inference
under weak identification]\label{prop:wald}

In the just-identified case \(d_z = 1\), write
\(u = \tilde Y - \tau_0 \tilde D\) for the structural score residual,
\(\Sigma = \mathbb{E}[\tilde Z^2 u^2]\), and \(V = \Sigma / g_{ZD}^2\)
for the Wald variance. (i) This \(V\) is \textbf{invariant to the scale
of} \(\tilde Z\): the magnitude of the reported standard error therefore
carries no evidence about instrument strength on its own, in either
direction. (ii) Under the local-to-zero sequence of \cite{staigerStock1997}, \(g_{ZD,n} = \gamma/\sqrt n\), the cross-fitted joint limit of
Lemma~\ref*{lem:cfjoint} (Supplementary \ref*{a.2-proposition-2-weak-identification-limit}) yields: \(\hat\tau\) is
inconsistent and \(O_p(1)\); the standard error neither vanishes nor explodes (\(\widehat{\mathrm{se}}\) and \(\widehat{\mathrm{se}}^{-1}\) are both \(O_p(1)\)); and the statistic whose variance is evaluated at
the \emph{hypothesized} value is asymptotically pivotal in absolute
value, its square being exactly the cross-fitted AR statistic of §\ref{estimation-and-inference}.
(iii) The \emph{feasible} Wald statistic
\((\hat\tau - \tau_0)/\widehat{\mathrm{se}}(\hat\tau)\), whose variance
evaluates residuals at \(\hat\tau\), is \textbf{non-pivotal} along the
sequence for every value of the score correlation
\(\rho = \mathrm{Corr}(\tilde Z u, \tilde Z \tilde D)\), including
\(\rho = 0\), and is standard normal only in the strong-identification
limit. Standard-normal critical values are therefore not generically
valid along the sequence, no fixed critical value controls size
uniformly over strength and score-correlation configurations, and the
nominal \(1 - \alpha\) Wald interval does not attain \(1-\alpha\)
coverage in general (Supplementary \ref*{a.2-proposition-2-weak-identification-limit}).

\end{proposition}

Proposition~\ref{prop:wald} shows why a near-zero estimate with a small
reported Wald standard error need not establish a precisely
estimated zero. Along the local-to-zero sequence, the feasible
Wald statistic has a nonstandard limiting distribution, so
conventional Wald confidence intervals need not attain nominal
coverage. This motivates the identification-robust inference
procedures described in Section~\ref{estimation-and-inference}.

\subsection{The anatomy of concentration
loss}\label{the-anatomy-of-concentration-loss}

Proposition~\ref{prop:wald} characterizes the consequences of weak
residualized identification for conventional Wald inference.
We now examine whether the choice of instrument residualization
can itself reduce population concentration. Holding the
conditioning set fixed and taking the treatment nuisance
to be correctly specified, we compare \(C(\pi)\) across
admissible residualizations. The following proposition
establishes that the true conditional mean
\(\pi_0(W)=\mathbb{E}[Z\mid W]\) maximizes this population
concentration index.

\begin{proposition}[Concentration is maximized at the true \(\pi_0\)]\label{prop:concentration}

Assume \(\mathbb{E}\lVert Z\rVert^2 + \mathbb{E}[D^2] < \infty\), that
\(\Omega(\pi_0)\) is positive definite, and that the relevance vector
\(g = \mathbb{E}[\tilde Z_0 \tilde D]\) is nonzero. With the treatment
nuisance correctly specified,
\(r = r_0 = \mathbb{E}[D\mid W]\), the residualized concentration
\(C(\pi) = g_{ZD}(\pi)'\,\Omega(\pi)^{-1} g_{ZD}(\pi)\) satisfies
\(C(\pi) \le C(\pi_0)\) for every square-integrable \(W\)-measurable
\(\pi\), with equality iff \(\pi = \pi_0\) when \(d_z = 1\), and for
\(d_z > 1\) iff the perturbation \(\pi_0 - \pi\) is
concentration-neutral (\(M\,\Omega(\pi_0)^{-1} g_{ZD} = 0\), with
\(M = \mathbb{E}[(\pi_0 - \pi)(\pi_0 - \pi)']\); Supplementary \ref*{a.3-proof-of-proposition-3}). The
prediction-optimal residualization \(\pi_0\) is therefore
concentration-optimal in population: a consistent flexible learner that
recovers a nonlinear \(\pi_0\) attains the \textbf{maximum}
concentration and cannot lower it.

\end{proposition}

The proof (Supplementary \ref*{a.3-proof-of-proposition-3}) closes only the population
channel: realized, finite-sample concentration can still fall, through
two diagnosable routes separated by the out-of-fold
predictive \(R^2\) of \(\hat\pi\) on \(Z\)
(\(R^2_{\text{oof}}(\hat\pi; Z)\)) and the residualized-variance ratio
\(\mathrm{Var}(\tilde Z)/\mathrm{Var}(Z)\), which are shown as follows.

\begin{diagnostic}[Routes by which the realized concentration can fall]
\leavevmode\par\smallskip
\begin{itemize}
\tightlist
\item
  \textbf{Overfit dilution}: a finite-sample
  \(\pi\)-\emph{estimation} pathology. \(\hat\pi\) has no
  out-of-fold predictive gain, \(R^2_{\text{oof}}(\hat\pi; Z) \le 0\),
  yet a large in-sample overfit gap; cross-fitting converts the overfit
  into idiosyncratic noise in the held-out residual, so
  \(\mathrm{Var}(\tilde Z)/\mathrm{Var}(Z) > 1\) and \(g_{ZD}\) is
  diluted. The dilution vanishes as \(n \to \infty\) and is removed by a
  restricted or cross-fit-stabilized \(\hat\pi\).
\item
  \textbf{Treatment-nuisance and conditioning effects}: operating
  through \(r\) and \(W\), not \(\pi\): (i) a richer set of controls
  \(W\) can lower the \emph{conditional} relevance
  \(\mathrm{Cov}(\tilde Z, \tilde D)\); a modeling choice, read
  off \(C\) (the Card ladder, §\ref{comparative-evidence-across-applications}); (ii) an \emph{under-specified}
  \(\hat r \neq \mathbb{E}[D\mid W]\) breaks the \(\pi\)-invariance of
  \(g_{ZD}\) that underlies Proposition \ref{prop:concentration} and can manufacture an
  apparent instrument ``absorption,'' which a correctly specified
  \(r\) removes (§\ref{nuisance-specification-and-residualized-strength}). Under stratified randomization of \(Z\)
  given \(W\) the assignment shares are known within strata, so
  \(\pi_0\) equals those shares; a learner can at most recover them,
  residualization removes only the across-strata variance of \(Z\),
  and \(C\) is preserved in population up to that design variation.
  In-fold noise fitting remains possible in principle; it is what the
  dilution diagnostics screen, and §\ref{comparative-evidence-across-applications} shows it is not materially
  present in the randomized cell.
\end{itemize}

\end{diagnostic}

\(R^2_{\text{oof}}(\hat\pi; Z) > 0\) with
\(\mathrm{Var}(\tilde Z)/\mathrm{Var}(Z) < 1\) means the learner is
recovering a predictable component of \(\pi_0\) (the linear baseline
judges whether it is nonlinear); under
correct nuisances,
this \emph{raises} \(C\) (Proposition \ref{prop:concentration}), so this signature alone is not a loss. A loss alongside it admits a
misspecified-\(r\) interaction, a change in conditioning, or
finite-sample noise; the diagnostics separate these only jointly. 
\section{Estimation and Inference}\label{estimation-and-inference}

We estimate the PLIV parameter using cross-fitted residualized moments
and use Anderson--Rubin (AR) confidence sets as the primary basis for
inference under potentially weak residualized identification.
Conventional Wald intervals provide a benchmark, while score-based
confidence sets provide a complementary procedure with multiple
instruments. We first describe estimation and the validity results
for AR and score inference under the stated conditions. We then
discuss survey weights and generated instruments. Section~\ref{design-a-structured-diagnostic-portfolio-in-residualized-strength}
sets out the common diagnostic and reporting protocol for the applications.

\subsection{Cross-fitted estimation}
\label{cross-fitted-residualized-estimation}

We implement the three-way protocol of Section~\ref{setup}, with
\(k(i)\) denoting observation \(i\)'s outer inference fold.
Within each outer training sample, candidate nuisance functions
are fitted and tuned on an estimation subsample, then selected
on a separate selection subsample using separate prediction
criteria for the outcome, treatment, and instrument nuisances.
The selected fits are evaluated on the inference fold without
refitting on the union of the estimation and selection subsamples.
A prespecified learner that undergoes no selection may instead
be fitted on the full outer training sample.
All fitted objects used for an observation are constructed
without using observations in its inference fold.

For clustered data, entire clusters are assigned to folds, and
all subsequent estimation and selection splits also partition
clusters. The assignment rule uses cluster sizes and, if needed,
randomization independent of the observed outcomes, treatments,
instruments, and controls. These rules preserve the separation
required by the clustered results below.

Write \(\hat\eta_{-k}=(\hat\mu_{-k},\hat r_{-k},\hat\pi_{-k})\)
for the nuisance estimates used to evaluate inference fold \(k\).
The estimated residuals are
\[
\hat{\tilde Y}_i=Y_i-\hat\mu_{-k(i)}(W_i),\qquad
\hat{\tilde D}_i=D_i-\hat r_{-k(i)}(W_i),\qquad
\hat{\tilde Z}_i=Z_i-\hat\pi_{-k(i)}(W_i).
\]
Define the pooled moments
\[
\hat g_{ZD}=\frac1n\sum_i\hat{\tilde Z}_i\hat{\tilde D}_i,\qquad
\hat g_{ZY}=\frac1n\sum_i\hat{\tilde Z}_i\hat{\tilde Y}_i,\qquad
\hat\Omega=\frac1n\sum_i\hat{\tilde Z}_i\hat{\tilde Z}_i'.
\]
The residualized-IV estimator is
\begin{equation}\label{eq:residualized-iv-estimator}
\hat\tau
=
\frac{\hat g_{ZD}'\hat\Omega^{-1}\hat g_{ZY}}
     {\hat g_{ZD}'\hat\Omega^{-1}\hat g_{ZD}}.
\end{equation}
For \(d_z=1\), this reduces to the ratio of the residualized
instrument--outcome and instrument--treatment moments; for
\(d_z>1\), it is the \(\hat\Omega^{-1}\)-weighted GMM estimator.
We report its heteroskedasticity- or cluster-robust sandwich
standard error as a conventional benchmark. When strong
identification cannot be maintained, Proposition~\ref{prop:wald} motivates
testing the null moment directly and inverting these tests
to construct confidence sets.

\subsection{Identification-robust inference}
\label{identification-robust-inference-is-the-primary-layer}

\subsubsection{Anderson--Rubin statistics}\label{subsec:ar-statistics}

For a candidate value \(t\), evaluate the cross-fitted null score
\[
\hat\psi_i(t)
=\hat{\tilde Z}_i
  \bigl(\hat{\tilde Y}_i-t\hat{\tilde D}_i\bigr).
\]
The covariance estimator is recomputed at each tested value.
With independent observations, the heteroskedasticity-robust AR
statistic is
\begin{equation}\label{eq:ar-iid}
\mathrm{AR}_n(t)
=
\left(\sum_i\hat\psi_i(t)\right)'
\left(\sum_i\hat\psi_i(t)\hat\psi_i(t)'\right)^{-1}
\left(\sum_i\hat\psi_i(t)\right).
\end{equation}
Unlike a Wald statistic, this construction evaluates the residual
at the hypothesized value and does not divide by an estimated
first-stage coefficient. Theorem~\ref*{thm:iid-import} of Supplementary \ref*{a.4-proof-of-theorem-1},
restated from \cite[Theorem~3]{cetin2026b}, establishes uniformly
correct asymptotic size using \(\chi^2_{d_z}\) critical values
under Assumptions A1--A7. We extend this result to clustered data.

For \(G\) i.i.d.\ clusters, define
\[
\hat S_c(t)=\sum_{i\in c}\hat\psi_i(t).
\]
The cluster-robust analogue is
\begin{equation}\label{eq:ar-clustered}
\mathrm{AR}_G(t)
=
\left(\sum_{c=1}^G\hat S_c(t)\right)'
\left(\sum_{c=1}^G\hat S_c(t)\hat S_c(t)'\right)^{-1}
\left(\sum_{c=1}^G\hat S_c(t)\right).
\end{equation}
The statistic aggregates scores within clusters before taking outer
products. Its validity also requires a cluster-level argument:
observation-level splitting need not separate training data from
dependent inference observations, and the nuisance-estimation
remainders must be controlled for cluster sums.

At the true value, the population score is mean zero regardless
of instrument strength. Orthogonality and the nuisance-rate
conditions control the effect of estimating the residuals,
while cluster aggregation accounts for within-cluster dependence.
The proof establishes equivalence to the oracle cluster scores,
a cluster-level central limit theorem, and covariance consistency.
Conditions C1--C8 in Supplementary \ref*{a.4-proof-of-theorem-1} require i.i.d.\ clusters
with bounded or suitably controlled random sizes, score moment
bounds and a uniformly nonsingular covariance, cluster-level
sample splitting, and uniform nuisance and product-rate conditions.
Within-cluster dependence is unrestricted subject to these bounds.

Condition C8 also requires exogeneity and stability of nuisance
conditional means given cluster size, together with conditional
moment and domination bounds. They ensure that the required centering
and error bounds remain valid when fold assignment depends on cluster
size. These conditions impose no positive lower bound on residualized
relevance. The following theorem establishes uniform asymptotic
validity under these conditions.

\begin{theorem}[Uniform validity of clustered cross-fitted AR inference]
\label{thm:clustered-ar}
Suppose Conditions C1--C8 in Supplementary \ref*{a.4-proof-of-theorem-1} hold, with
\(G\to\infty\) and fixed \(d_z\). For any fixed
\(\alpha\in(0,1)\),
\[
\sup_{P\in\mathcal P_G}
\left|
P\!\left\{\mathrm{AR}_G(\tau_0)>
\chi^2_{d_z,1-\alpha}\right\}-\alpha
\right|
\longrightarrow 0.
\]
Consequently, the inverted set
\[
\mathcal C_{\mathrm{AR},G}(1-\alpha)
=
\left\{t\in\mathbb R:
\mathrm{AR}_G(t)\le\chi^2_{d_z,1-\alpha}\right\}
\]
has coverage tending to \(1-\alpha\) uniformly over
\(\mathcal P_G\).
\end{theorem}

The class \(\mathcal P_G\) includes strong-identification laws
and relevance sequences of order \(G^{-1/2}\). The same critical
values therefore apply across these identification regimes,
subject to the stated conditions. The proof is in Supplementary \ref*{a.4-proof-of-theorem-1}.

\subsubsection{Score inference with multiple instruments}\label{subsec:score-inference}

The AR statistic evaluates the full \(d_z\)-dimensional vector
of null moments. A complementary score statistic projects these
moments onto an adjusted Jacobian direction. The two procedures
coincide with a single instrument, so the score test provides
a distinct inferential option in over-identified models.
Because \(\tau\) is scalar, the score test has a one-degree-of-freedom
reference under the conditions stated below.

Define the observation-level Jacobian contribution
\[
\hat q_i
=\partial_t\hat\psi_i(t)
=-\hat{\tilde Z}_i\hat{\tilde D}_i.
\]
To cover both sampling schemes, let \(N\) be the number of
independent sampling units. For i.i.d.\ observations,
\(N=n\), \(\hat S_a(t)=\hat\psi_a(t)\), and
\(\hat Q_a=\hat q_a\). For clustered observations,
\(N=G\), \(\hat S_a(t)\) is the cluster sum defined above,
and \(\hat Q_a=\sum_{i\in a}\hat q_i\).
Set
\begin{align*}
\hat m_N(t)&=\frac1N\sum_{a=1}^N\hat S_a(t),
&
\hat g_N&=\frac1N\sum_{a=1}^N\hat Q_a,\\
\hat V_{ff}(t)&=\frac1N\sum_{a=1}^N
                  \hat S_a(t)\hat S_a(t)',
&
\hat V_{\theta f}(t)&=\frac1N\sum_{a=1}^N
                  (\hat Q_a-\hat g_N)\hat S_a(t)'.
\end{align*}
The adjusted Jacobian and the cross-fitted score statistic are
\begin{align}
\hat D_N(t)
&=\hat g_N-\hat V_{\theta f}(t)
                 \hat V_{ff}(t)^{-1}\hat m_N(t),
\label{eq:adjusted-jacobian}\\
\mathrm K_N(t)
&=
\frac{N\bigl[\hat m_N(t)'\hat V_{ff}(t)^{-1}
                         \hat D_N(t)\bigr]^2}
     {\hat D_N(t)'\hat V_{ff}(t)^{-1}\hat D_N(t)}.
\label{eq:score-statistic}
\end{align}
This is the scalar-parameter score construction of \cite{kleibergen2005}, applied to the cross-fitted moments.

AR validity alone does not establish the joint approximation required
by the score statistic. We therefore impose two additional conditions
in Supplementary \ref*{a.5-proof-of-theorem-2}. K1 imposes moment bounds on the stacked score
and Jacobian, conditional second-moment bounds, and non-degeneracy of
the Jacobian covariance after projection on the score. K3 requires
both covariance blocks to use the same sampling-unit aggregation:
with clustered data, both the score and the Jacobian are summed
within clusters before the covariance blocks are formed.
These conditions make the adjusted direction well defined
asymptotically even when population relevance approaches zero.

\begin{theorem}[Uniform validity of the cross-fitted score test]
\label{thm:score}
Assume A1--A7 in the i.i.d.\ case or C1--C8 in the clustered
case, together with K1 and K3 in Supplementary \ref*{a.5-proof-of-theorem-2}.
Let \(N=n\) in the i.i.d.\ case and \(N=G\) in the clustered
case, with all score and Jacobian contributions aggregated
at the corresponding sampling-unit level. For fixed \(d_z\)
and any fixed \(\alpha\in(0,1)\),
\[
\sup_{P\in\mathcal P_N}
\left|
P\!\left\{\mathrm K_N(\tau_0)>
\chi^2_{1,1-\alpha}\right\}-\alpha
\right|
\longrightarrow 0,
\]
where \(\mathcal P_N=\mathcal P_n\) or \(\mathcal P_G\)
according to the sampling scheme.
\end{theorem}

Inverting the score test gives the confidence set
\[
\mathcal C_{\mathrm K,N}(1-\alpha)
=\{t\in\mathbb R:\mathrm K_N(t)\le\chi^2_{1,1-\alpha}\},
\]
with uniform asymptotic coverage under the same conditions.
The proof is in Supplementary \ref*{a.5-proof-of-theorem-2}.
When \(d_z=1\), \(\mathrm K_N(t)=\mathrm{AR}_N(t)\), using
the continuous extension if the adjusted Jacobian vanishes.
For multiple instruments, the theorem establishes validity
but does not assert a uniform power advantage over AR.

\subsection{Survey weights and generated instruments}
\label{survey-weights-and-generated-instruments}

Survey weights change both the score and, in general, the
nuisance targets. Weighting an existing unweighted score need not
preserve its centering and orthogonality; the conditional means and
exclusion restriction must correspond to the weighted target law.
For known weights satisfying
\(0<c_\omega\le\omega_i\le C_\omega<\infty\),
let \(P_\omega\) satisfy
\(dP_\omega/dP=\omega/\mathbb E_P[\omega]\).
If \(\omega\) is a function of \(W\) alone, the weighted and
unweighted conditional means coincide. Otherwise, nuisance
fitting and selection use the weighted prediction criterion.

Estimate the conditional means under \(P_\omega\) and form
the corresponding cross-fitted residuals. Replace the score
and Jacobian contributions by
\[
\hat\psi_i^\omega(t)
=\omega_i\hat{\tilde Z}_i^\omega
 \bigl(\hat{\tilde Y}_i^\omega-t\hat{\tilde D}_i^\omega\bigr),
\qquad
\hat q_i^\omega
=-\omega_i\hat{\tilde Z}_i^\omega\hat{\tilde D}_i^\omega.
\]
The AR and score formulas then apply to these contributions,
with weighted contributions summed within clusters when
appropriate. Weighted moments also replace the moments in
\eqref{eq:residualized-iv-estimator} for point estimation.

Proposition~\ref*{prop:weighted} in Supplementary \ref*{a.6-weighted-and-generated-instrument-extensions} establishes validity
under the tested-value exclusion condition W-ID\((t)\),
the appropriate nuisance-rate conditions, and the required
weighted-score moment and non-degeneracy conditions; score inference
also requires K1 and K3 for the weighted contributions. In the clustered case, the moment and covariance conditions are
imposed on the weighted cluster-score sums.

When the instrument is itself estimated, construction error enters
its residual in addition to the error in its conditional mean.
The observed-instrument argument does not control this extra term.
Construction must also be nested within the folds to avoid using
inference observations to generate their own instruments.
For example, a leniency
instrument for an inference observation is constructed
using treatment data outside that observation's inference
cluster fold, and its subsequent residualization follows the same
fold-exclusion rule. This makes the constructed instrument function
measurable with respect to the training data. Instrument estimation
is then treated as an additional nuisance-estimation step.
Proposition~\ref*{prop:generated} in Supplementary \ref*{a.6-weighted-and-generated-instrument-extensions} extends
Theorem~\ref{thm:clustered-ar} under the generated instrument's
exclusion condition, a construction-error rate, and an augmented
product-rate condition. These rates control the additional score
remainder. Through its identity with AR, score inference is also
covered when \(d_z=1\).

\section{Monte Carlo Studies}\label{monte-carlo-studies}

The experiments assess the coverage and informativeness of
identification-robust confidence sets and the role of nuisance
specification in residualized strength. Supplementary experiments on
identification-aware learner selection and selection--inference overlap
are reported in Supplementary Table~\ref*{tab:b3} and Supplementary~\ref*{app:e}.

\subsection{Simulation design}\label{simulation-design}

The core design is a partially linear IV model with correlated Gaussian
controls, \(W\sim N(0,\Sigma_W)\), where
\((\Sigma_W)_{ik}=0.5^{|i-k|}\), and
\[
\begin{aligned}
Z_j &= \sqrt{\rho}\,b_j(W)+\sqrt{1-\rho}\,e_{Z,j},\\
D &= W'\beta_r+f_s\sum_{j=1}^{d_z}Z_j+\kappa A+\nu,\\
Y &= \tau_0D+W'\beta_g+\kappa A+\zeta.
\end{aligned}
\]
The innovations \(e_{Z,j}\), \(A\), \(\nu\), and \(\zeta\) are mutually
independent standard normals, independent of \(W\). The shared
confounder \(A\), with \(\kappa=0.8\), creates endogeneity, while
\(\tau_0=1\) is fixed. The functions \(b_j\) are zero, linear, or
nonlinear, with fixed population standardization. Thus
\(\pi_{0,j}=\sqrt{\rho}\,b_j\) and the treatment conditional mean is
\(r_0(W)=W'\beta_r+f_s\sum_j\pi_{0,j}(W)\).

The core sweeps vary \(f_s\) and, where specified, \(\rho\) to span
effective-\(F\) values from about \(1\) to \(500\) at a fixed target.
Estimated nuisances use five cross-fitting folds. Additional arms
examine nonlinear structural functions, clustering, survey weighting,
generated instruments with nested selection, and over-identification
up to \(d_z=30\). The weighted arm targets \(\tau_w=1.2\), distinct
from the unweighted target \(1.5\); clustered arms use cluster-aware
folds and variances. Supplementary \ref*{app:c} gives the designs, sample
sizes, replication counts, and seeds; result tables report Monte
Carlo standard errors.

\subsection{Cross-fitted AR coverage and power}
\label{a-monte-carlo-coverage-evaluation-for-the-cross-fitted-ar-set}

\begin{figure}[tbp]
\centering
\includegraphics[width=\linewidth]{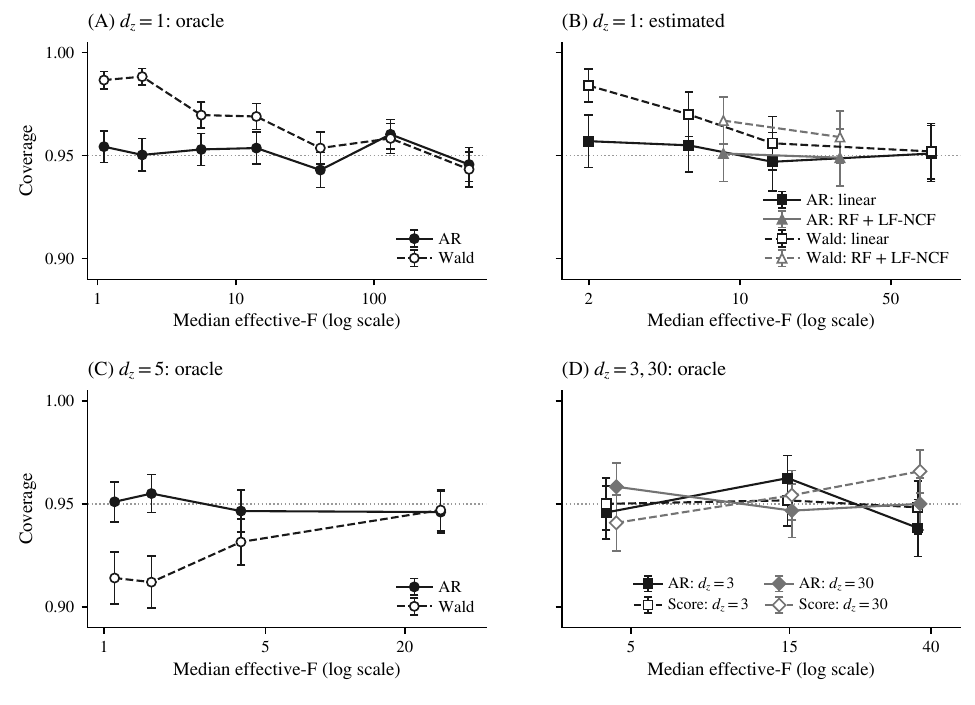}
\caption{Monte Carlo coverage across identification regimes:
AR and Wald in (A)--(C), AR and score in (D).
(A) Single-instrument oracle nuisances.
(B) Estimated-linear nuisances and a nonlinear design with random-forest
outcome and treatment nuisances and nested predictive selection of
instrument residualization.
(C) Five instruments with oracle nuisances.
(D) Three or thirty instruments with oracle nuisances.
Bars are \(\pm2\) point-specific Monte Carlo standard errors;
the horizontal line is nominal \(0.95\). Stress and constructed-instrument
shortfalls are reported in the text and Supplementary Tables \ref*{tab:b2} and \ref*{tab:b7}.}
\label{fig:mc-coverage}
\label{fig:fig2}
\end{figure}

Figure~\ref{fig:mc-coverage} compares cross-fitted Anderson--Rubin (AR),
Wald, and score coverage across oracle, estimated-linear, nonlinear,
and over-identified settings, each with at least \(1,000\)
replications. Across these and the additional nonlinear-nuisance settings
in Supplementary Table \ref*{tab:b2}, AR coverage ranges from \(0.938\) to \(0.968\), against
nominal \(0.95\). At the two calibrated flexible-nuisance points,
with median effective-\(F\) of \(8.4\) and \(29.1\), coverage is
\(0.951\) and \(0.949\); Wald coverage is \(0.967\) and
\(0.959\). Under weak just-identification, high Wald coverage can
reflect unstable estimates and inflated standard errors. Under over-identification,
Wald instead undercovers, reaching \(0.91\) at \(d_z=5\) and
\(0.80\) at \(d_z=10\) (Supplementary Tables \ref*{tab:b2} and \ref*{tab:b7}).

Dependence and weighting matter for the comparison. Cluster-robust AR
covers at \(0.948\) and \(0.951\), versus \(0.51\) and \(0.69\)
when clustering is ignored. Weighted AR covers \(\tau_w\) at
\(0.946\) and \(0.947\); unweighted inference, centered on a different
target, covers it at only \(0.59\) and \(0.09\) (Supplementary Table \ref*{tab:b2}).

The stress experiments reveal finite-sample coverage shortfalls.
Across eleven adversarial points, AR coverage ranges from \(0.920\)
to \(0.961\). The minimum occurs with flexible instrument
residualization at effective-\(F\approx33.5\)
(Supplementary Table \ref*{tab:b7}). In the combined clustered, generated-instrument, and
nested-selection arm, the weak constructed-instrument row covers at
\(0.912\). Its
observed-instrument control covers at \(0.956\) at the same design
strength, but construction lowers median effective-\(F\) from
\(9.6\) to \(2.0\); a control matched on attained strength covers
at \(0.936\). At moderate strength, constructed and observed
instruments cover at \(0.944\) and \(0.952\); increasing construction
support from \(m=6\) to \(m=20\) cases per cluster leaves constructed
coverage at \(0.944\) (Supplementary Table \ref*{tab:b2}). 

Supplementary Table~\ref*{tab:b9} confirms that test-based and extracted-set coverage agree
in every reported setting. In the single-instrument sweeps, stronger
identification yields narrower finite-length sets and higher rejection
rates at \(\tau_0+0.5\). In the structure-matched Angrist--Krueger
arm with \(51\) clusters, however, AR sets are almost always the whole
line at \(d_z=30\): conservative coverage comes with little
informativeness (Supplementary Table~\ref*{tab:b10}).

\subsection{Nuisance specification and residualized strength}
\label{nuisance-specification-and-residualized-strength}

\begin{figure}[tbp]
\centering
\includegraphics[width=0.92\linewidth]{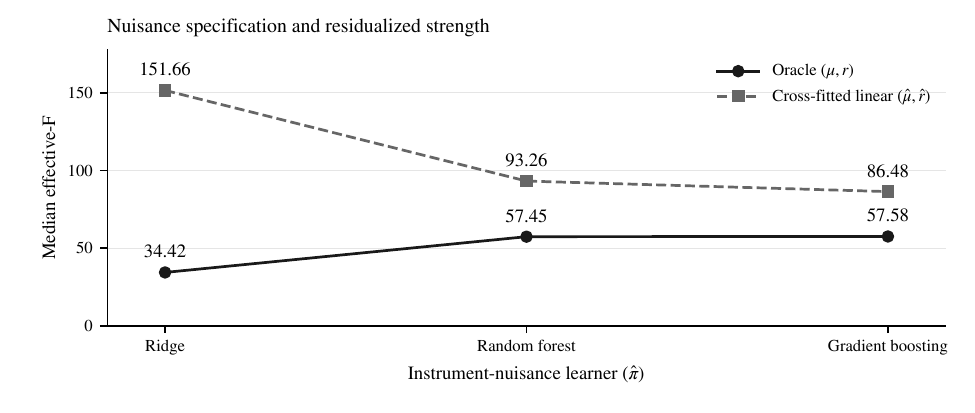}
\caption{Median residualized effective-\(F\) across ridge, random-forest,
and gradient-boosting instrument residualizations. The two specifications use oracle outcome and treatment
nuisances or cross-fitted linear outcome and treatment nuisances.
Both nuisances change; the strength comparison concerns the treatment
residual. Supplementary \ref*{app:c} and \ref*{app:d1} give the design and diagnostic detail.}
\label{fig:mc-nuisance-strength}
\end{figure}

A separate experiment examines how nuisance specification affects the
relationship between instrument prediction and residualized strength.
With nonlinear \(\pi_0\),
\(n=1{,}500\), and \(100\) replications, the random forest predicts
\(Z\) better out of fold than a linear learner
(\(R^2=0.50\) versus \(0.18\)) and leaves less residualized variance
(ratios \(0.50\) versus \(0.82\)). The strength ranking nevertheless
depends on the treatment nuisance (Figure~\ref{fig:mc-nuisance-strength}).
Under correct oracle outcome and treatment nuisances, the forest
attains effective-\(F\) about \(57\), versus \(34\) for ridge
residualization, consistent with Proposition~\ref{prop:concentration}'s population benchmark.
With cross-fitted linear outcome and treatment nuisances, that ordering
reverses: about \(93\) versus \(152\).

The true treatment conditional mean contains the nonlinear component
\(f_s\pi_0(W)\). A linear \(\hat r\) does not fully remove this component,
leaving structure in the treatment residual that is correlated with
predictable instrument variation. A flexible \(\hat\pi\) removes more of
this predictable variation from the instrument, leading to a
decline in effective-\(F\) in this design. Thus, when random-forest
instrument residualization yields a lower effective-\(F\), the treatment
nuisance specification should be checked first.

\section{Applications}\label{empirical-applications}

\subsection{Empirical design and diagnostic protocol}
\label{design-a-structured-diagnostic-portfolio-in-residualized-strength}

\begin{figure}[tbp]
\centering
\begin{adjustbox}{max width=\textwidth}
\begingroup
\def\baselinestretch{1}\selectfont
\begin{tikzpicture}[
  text=black,font=\rmfamily\fontsize{10}{12.4}\selectfont,
  flow/.style={-{Latex[length=1.8mm,width=1.2mm]},
    draw=black,line width=0.5pt},
  bus/.style={draw=black,line width=0.5pt},
  stage/.style={rectangle,draw=black,fill=white,
    line width=0.5pt,text width=15.4cm,align=center,
    inner xsep=3mm,inner ysep=3mm,anchor=north},
  lane/.style={rectangle,draw=black,fill=white,
    line width=0.5pt,inner xsep=3mm,inner ysep=3mm,
    anchor=north,align=left}
]

\node[stage] (results) at (0,0) {
  {\bfseries 1\quad Cross-fitted results}\\[4pt]
  Out-of-fold predictions and residuals\\[2pt]
  Cluster-aware folds \quad\textbullet\quad Separate \(\mu,r,\pi\)
};
\coordinate (branchTop) at ($(results.south)+(0,-0.80)$);

\node[lane] (diagnosis) at ($(branchTop)+(-4.15,0)$) {
  \begin{minipage}[t][8.15cm][t]{7.1cm}
  \centering
  {\bfseries 2a\quad Source diagnosis}\par\vspace{7pt}
  \(R^2_{\mathrm{oof}}(\mu,r,\pi),\ \Delta_\pi,\ V_\pi\)\\[3pt]
  Calibration, \(F_{\mathrm{res}},\ \beta_{\mathrm{res}}\)
  \par\vspace{7pt}\hrule height 0.3pt\vspace{7pt}
  
  \par\vspace{9pt}
  \raggedright
  \textbf{Leakage / misfit}\\[2pt]
  High \(R^2_{\mathrm{oof}}(\mu,r,\pi)\) or disagreement\\
  weakens after fold / nuisance correction
  \par\vspace{9pt}
  \textbf{Overfit dilution}\\[2pt]
  \(R^2_{\mathrm{oof},\pi}\leq0,\ \Delta_\pi\text{ large},\ V_\pi>1;\
    F_{\mathrm{res}}\downarrow\)
  \par\vspace{9pt}
  \textbf{Conditioning effect}\\[2pt]
  Fixed sample and \(Z\); richer \(W\)\\
  \(R^2_{\mathrm{oof},\pi},V_\pi\text{ stable};\
    \beta_{\mathrm{res}}\downarrow,\ F_{\mathrm{res}}\downarrow\)
  \par\vspace{9pt}
  \textbf{Stable pattern}\\[2pt]
  Diagnostics stable across learners
  \end{minipage}
};

\node[lane] (inference) at ($(branchTop)+(4.15,0)$) {
  \begin{minipage}[t][8.15cm][t]{7.1cm}
  \centering
  {\bfseries 2b\quad Effect inference}\par\vspace{7pt}
  AR confidence set\\[3pt]
  At all strength levels
  \par\vspace{7pt}\hrule height 0.3pt\vspace{7pt}
 
  \par\vspace{15pt}
  \raggedright
  \textbf{Effect values}\\[3pt]
  Compatible values; inclusion of zero
  \par\vspace{22pt}
  \textbf{Informativeness}\\[3pt]
  Set width and unboundedness
  \par\vspace{22pt}
  \textbf{Sensitivity}\\[3pt]
  Learner choice and sample splitting
  \end{minipage}
};

\coordinate (split) at ($(results.south)+(0,-0.35)$);
\draw[bus] (results.south) -- (split);
\draw[bus] (split -| diagnosis.north) -- (split -| inference.north);
\draw[flow] (split -| diagnosis.north) -- (diagnosis.north);
\draw[flow] (split -| inference.north) -- (inference.north);

\coordinate (merge) at ($(diagnosis.south)+(4.15,-0.40)$);
\draw[bus] (diagnosis.south) -- (diagnosis.south |- merge)
  -- (inference.south |- merge) -- (inference.south);
\node[stage] (interpretation) at ($(merge)+(0,-0.40)$) {
  {\bfseries 3\quad Integrated interpretation}\\[4pt]
  Effect conclusions, informed by diagnostic patterns\\[2pt]
  Possible sources of weakness or sensitivity \quad Limits of the evidence
};
\draw[flow] (merge) -- (interpretation.north);

\end{tikzpicture}
\endgroup
\end{adjustbox}
\caption{The common diagnostic and inferential workflow.
\(\Delta_\pi=R^2_{\mathrm{in},\pi}-R^2_{\mathrm{oof},\pi}\),
\(V_\pi=\operatorname{Var}(\widetilde Z)/\operatorname{Var}(Z)\),
and \(\beta_{\mathrm{res}}\) is the single-instrument first-stage slope.
Diagnostic patterns are interpreted jointly after checking folds and
nuisance specification.}
\label{fig:section5-diagnostic-workflow}
\end{figure}
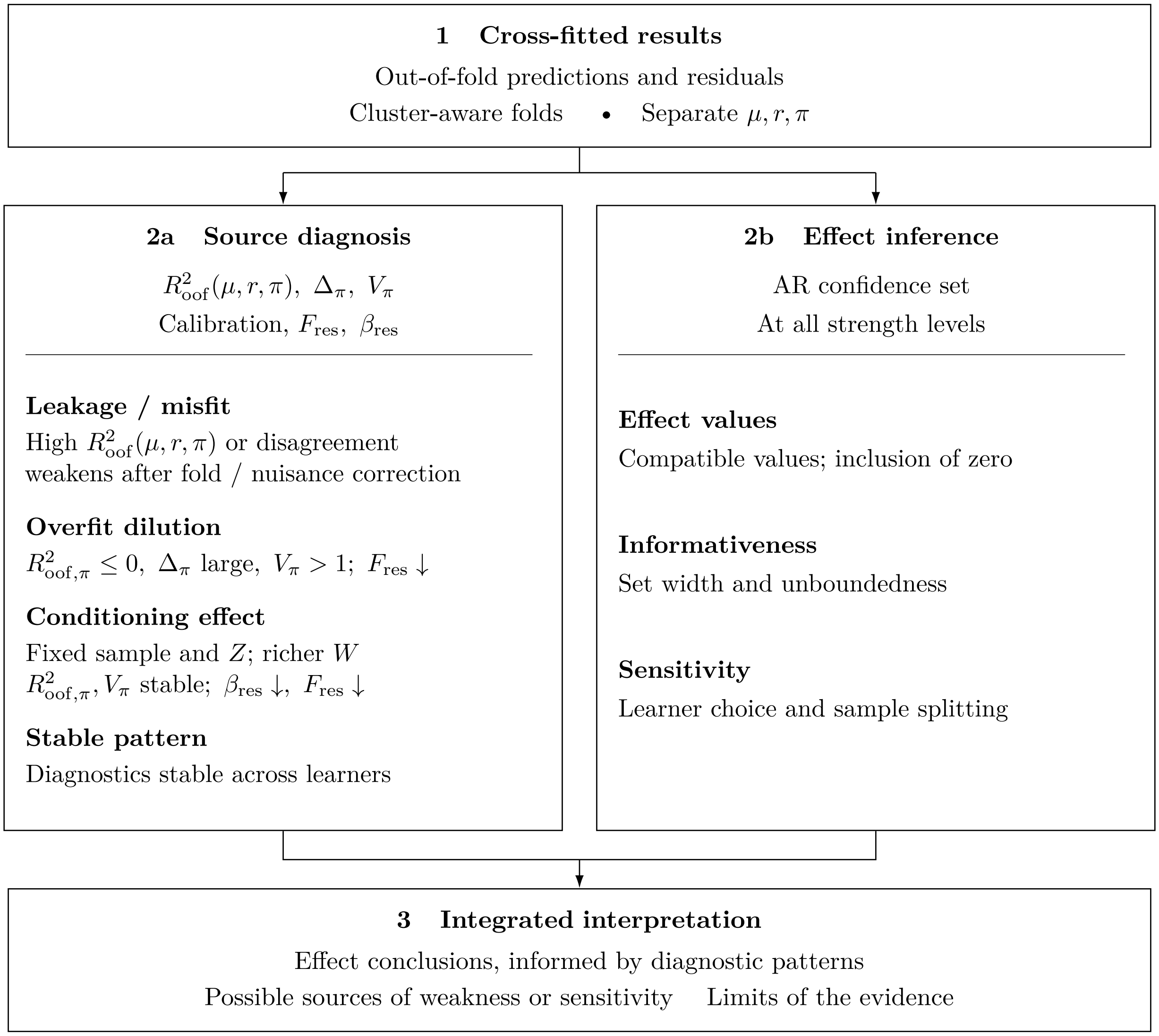

Five applications examine the sources of residualized-strength changes
and their consequences for effect inference. Figure~\ref{fig:section5-diagnostic-workflow}
summarizes the common protocol. We fit outcome, treatment, and instrument nuisances separately using
out-of-fold predictions and folds that respect dependence. Instrument
fit, calibration, the overfit gap \(\Delta_\pi\), and the residualized
variance ratio \(V_\pi\) are read jointly with the first-stage slope
and strength. The AR set remains primary at every strength level;
learner and split comparisons assess sensitivity. Table~\ref{tab:3} reports estimates and inference;
Table~\ref{tab:4} states each application's target, assumptions, and
theoretical support. Descriptive statistics and diagnostics are
in Supplementary Tables \ref*{tab:1}--\ref*{tab:2} and \ref*{tab:b1}. Set extraction, including
unbounded or empty sets and numerical truncation, follows Supplementary \ref*{a.4-proof-of-theorem-1}.

Strength is the robust residualized first-stage Wald statistic divided
by the numerical rank of the residualized instrument covariance. It equals the \cite{montielOleaPflueger2013} effective \(F\) for \(d_z=1\). With multiple
instruments, it serves only as a strength diagnostic. Over-identified
models add score sets under Section~\ref{identification-robust-inference-is-the-primary-layer}'s conditions,
\(J_{\mathrm{CUE}}\), and supplementary conditional QLR.

\clearpage
\begin{table}[p]
\centering
\small
\caption{Main results: estimates and identification-robust inference}
\label{tab:3}
\begin{adjustbox}{max width=\linewidth}
\begin{tabular}{llrrrlr}
\toprule
Cell & Estimator & $\hat\tau$ & Wald SE & eff.-$F$ & AR 95\% set & $N$ \\
\midrule
OHIE & OLS & $1.196$ & $0.063$ & — & — & 23,441 \\
 & 2SLS & $1.034$ & $0.173$ & 2098 & — & 23,441 \\
 & DML (ridge) & $1.037$ & $0.174$ & 2093 & $[0.70, 1.38]$ & 23,441 \\
 & DML (RF) & $1.050$ & $0.175$ & 2024 & $[0.71, 1.39]$ & 23,441 \\
 & DML (GB) & $1.041$ & $0.172$ & 2094 & $[0.70, 1.38]$ & 23,441 \\
Examiner & OLS & $-0.158$ & $0.005$ & — & — & 91,282 \\
 & 2SLS & $-0.497$ & $0.189$ & 24.4 & — & 91,282 \\
 & DML (ridge) & $-0.480$ & $0.223$ & 20.1 & $[-1.02, -0.04]$ & 91,282 \\
 & DML (RF) & $-0.439$ & $0.263$ & 14.2 & $[-1.12, 0.10]$ & 91,282 \\
 & DML (GB) & $-0.455$ & $0.219$ & 20.6 & $[-0.98, -0.02]$ & 91,282 \\
CollegeDistance & OLS & $-0.001$ & $0.001$ & — & — & 4,739 \\
 & 2SLS & $0.052$ & $0.020$ & 9.6 & — & 4,739 \\
 & DML (ridge) & $-0.003$ & $0.113$ & 14.8 & $[-0.227, 0.766]$ & 4,739 \\
 & DML (RF) & $-0.006$ & $0.106$ & 7.9 & $[-0.263, 0.346]$ & 4,739 \\
 & DML (GB) & $0.001$ & $0.118$ & 8.9 & $[-0.287, 0.381]$ & 4,739 \\
Card & OLS & $0.075$ & $0.004$ & — & — & 3,010 \\
 & 2SLS & $0.132$ & $0.054$ & 13.5 & — & 3,010 \\
 & DML (ridge) & $0.127$ & $0.052$ & 15.9 & $[0.03, 0.27]$ & 3,010 \\
 & DML (RF) & $0.152$ & $0.054$ & 15.2 & $[0.05, 0.31]$ & 3,010 \\
 & DML (GB) & $0.118$ & $0.049$ & 18.2 & $[0.02, 0.24]$ & 3,010 \\
\bottomrule
\end{tabular}
\end{adjustbox}
\par\smallskip
\begin{minipage}{\linewidth}
\footnotesize \emph{Notes.} Wald standard errors are clustered by household (OHIE), court-by-date shift (examiner), or state proxy (CollegeDistance DML); Card and the CollegeDistance OLS/2SLS rows use heteroskedasticity-robust errors. OHIE uses survey weights. All reported effective-$F$ values use the shared cross-fitted first-stage diagnostic, equal to the Montiel Olea--Pflueger effective $F$ at $d_z=1$. AR sets are primary at every strength level (Section~\ref{identification-robust-inference-is-the-primary-layer}). Ridge is cross-validated and closely matches the linear benchmark; redundant linear rows are omitted. Supplementary Table \ref*{tab:b1} reports learner selections and nuisance diagnostics; the Angrist--Krueger results are in Supplementary Table \ref*{tab:b6} and Section~\ref{comparative-evidence-across-applications}. The CollegeDistance OLS/2SLS errors are not directly comparable to its clustered DML errors: using the same $43$ groups, conventional 2SLS has SE $0.0646$, AR set $[-0.096, 0.234]$, and first-stage $F=11.75$.
\end{minipage}
\end{table}

\clearpage

\begin{landscape}
\begin{table}[p]
\centering
\footnotesize
\caption{Statistical and causal targets by application}
\label{tab:4}
\begin{adjustbox}{max width=\linewidth}
\begin{tabular}{>{\raggedright\arraybackslash}p{0.07\linewidth} >{\raggedright\arraybackslash}p{0.16\linewidth} >{\raggedright\arraybackslash}p{0.21\linewidth} >{\raggedright\arraybackslash}p{0.205\linewidth} >{\raggedright\arraybackslash}p{0.08\linewidth} >{\raggedright\arraybackslash}p{0.265\linewidth}}
\toprule
Cell & Estimand and moment & Causal assumptions (source) & Target population and weights & Dependence; $d_z$ & Evidence layer \\
\midrule
OHIE & weighted PLIV coefficient $\tau_w$ on ever-Medicaid (binary $D$); survey-weighted orthogonal score & lottery randomization within household-size strata; exclusion: selection affects visits only through coverage \citep{finkelstein2012} & survey-weighted twelve-month responders (weights re-weight toward the eligible population); response-selection sensitivity via Lee trimming (Supplementary Table \ref*{tab:b5}) & household clusters; $d_z=1$ & Prop.~\ref*{prop:weighted} (regime (b)) with Thm.~\ref{thm:clustered-ar} (household clusters); clustered and survey-weighted §\ref{a-monte-carlo-coverage-evaluation-for-the-cross-fitted-ar-set} arms \\
Examiner & PLIV coefficient on pretrial release (binary $D$); generated fold-nested cluster-jackknife leniency instrument & quasi-random assignment within court-by-time, balance per \cite{chyn2025}; exclusion through release & Miami--Dade cases of judges with at least $200$ cases (source-study restriction) & court-by-date shift clusters; $d_z=1$ (generated) & Thm.~\ref{thm:clustered-ar} via Prop.~\ref*{prop:generated} (fold-nested construction); combined §\ref{a-monte-carlo-coverage-evaluation-for-the-cross-fitted-ar-set} arm (Supplementary Table \ref*{tab:b2}) \\
College-Distance & diagnostic stress test, not a clean causal cell; PLIV coefficient on years of education (multivalued $D$) & conditional-on-observables only; region-determined outcome undermines a clean exclusion reading (Supplementary \ref*{app:d4}) & complete cases with positive wage; folds on the $43$-group (region, tuition) state proxy & state-proxy groups; $d_z=1$ & Thm.~\ref{thm:clustered-ar} ($d_z=1$, $43$ state-proxy groups; modest $G$); clustered §\ref{a-monte-carlo-coverage-evaluation-for-the-cross-fitted-ar-set} arm; §\ref{comparative-evidence-across-applications} leakage diagnostics \\
Card & PLIV coefficient on years of education (multivalued $D$) & conditional independence of college proximity given controls; maintained, not testable at $d_z=1$ & full NLS young-men complete-case sample ($N=3{,}010$); ladder on the $2{,}182$-man background-complete subsample & heterosked.-robust; $d_z=1$ & Thm.~\ref*{thm:iid-import} \citep[imported;][]{cetin2026b}; i.i.d.\ §\ref{a-monte-carlo-coverage-evaluation-for-the-cross-fitted-ar-set} arms \\
Angrist--Krueger & PLIV coefficient on years of schooling (multivalued $D$) & quarter-of-birth exclusion and independence as maintained in \cite{angristKrueger1991}; many-weak caveats of \cite{boundJaegerBaker1995} & 1980 Census men born 1930--39 & state-of-birth clusters ($51$); $d_z=3, 30$ & Thm.~\ref{thm:clustered-ar} (AR) and Thm.~\ref{thm:score} (score; K1 non-degeneracy), via the within-transformation bridge Lem.~\ref*{lem:within-bridge}, with the $G=51$ boundary caveat (Supplementary \ref*{app:d6}); clustered QLR reported as a construction; i.i.d.\ over-identified §\ref{a-monte-carlo-coverage-evaluation-for-the-cross-fitted-ar-set} arms at matched counts and strengths \\
\bottomrule
\end{tabular}
\end{adjustbox}
\par\vspace{1pt}
\begin{minipage}{\linewidth}
\footnotesize \emph{Notes.} The maintained target is the homogeneous-coefficient PLIV parameter (Section~\ref{setup}). Under heterogeneity, multivalued schooling treatments would instead yield an average-causal-response-type weighted average; a complier interpretation for binary treatment would additionally require monotonicity, which is not imposed here. The evidence column pairs the applicable asymptotic result with simulation designs in Section~\ref{a-monte-carlo-coverage-evaluation-for-the-cross-fitted-ar-set}. Those designs assess finite-sample performance under specified dependence structures, not IV validity in the applications. Clustered conditional QLR is exploratory and carries no established reference distribution here.
\end{minipage}
\end{table}
\end{landscape}

\subsection{Comparative evidence across applications}
\label{comparative-evidence-across-applications}

OHIE and baseline Card show learner stability at different strength
levels. In OHIE \citep{finkelstein2012}, lottery selection
instruments Medicaid enrollment for twelve-month primary-care visits.
Across learners, effective-\(F\) stays around \(2000\)--\(2100\),
\(R^2_{\mathrm{oof},\pi}\approx0\), and \(V_\pi\approx1\).
Estimates remain near \(1.04\) additional visits, with AR sets close
to \([0.70,1.39]\). \cite{card1995} instruments schooling with college
proximity in a log-wage equation. On \(3{,}010\) men, estimates range
from \(0.118\) to \(0.152\), effective-\(F\) from \(15.2\) to
\(18.2\), and bounded AR sets overlap.

OHIE targets the PLIV coefficient under the survey-weighted law of
\(23{,}441\) complete-case responders. Weights enter the entire
pipeline, with household-clustered inference under Proposition~\ref*{prop:weighted}'s
regime (b) and weighted-null conditions. The diagnostics support
approximately constant instrument propensity among responders;
lottery randomization alone fixes it within household-size strata
in the lottery population. Response selection still matters:
scaling the Lee ITT bounds by the untrimmed first stage gives
sensitivity ranges \([0.77,2.05]\) with survey weights and
\([-4.0,4.3]\) without them (Supplementary Table \ref*{tab:b5}).

Examiner leniency instead exposes inference sensitivity despite
similar point estimates. Miami--Dade pretrial data \citep{chyn2025} use judge leniency to instrument release in a
conviction equation. The \(91{,}282\)-case sample retains judges
with at least \(200\) cases. The leave-one-shift-out instrument is
nested within training folds, with court-by-date shift clustering. The examiner random forest fits poorly out of fold despite substantial
in-sample fit: \(R^2_{\mathrm{oof},\pi}<0\), \(V_\pi=1.26\), and
effective-\(F\) falls from ridge's \(20.1\) to \(14.2\), consistent
with overfit dilution. The estimate shifts from \(-0.480\) to
\(-0.439\), but the AR set widens from \([-1.02,-0.04]\) to
\([-1.12,0.10]\). With all three nuisances fixed to ridge, the AR set includes zero in two of six sample partitions (Supplementary Table \ref*{tab:b12}). Thus the negative estimates suggest that release lowers
conviction under the maintained assumptions, while exclusion of zero
depends on the learner and split.

Within Card, conditioning changes strength even when instrument fit
is stable. Holding college proximity and the \(2{,}182\)-man sample
fixed, a five-rung control ladder lowers effective-\(F\) from about
\(7.5\) to \(2.5\), non-monotonically. Instrument fit and residualized
variance barely change, but the first-stage slope falls from \(0.26\)
to \(0.12\): the loss concerns treatment--instrument covariance.
Linear-\(\pi\) estimates stay around \(0.04\)--\(0.08\) (Supplementary \ref*{app:d5}), while AR
sets become the whole line at the three richest rungs (Supplementary Table \ref*{tab:b4}).
The ladder diagnoses concentration, not competing causal specifications:
upper rungs include contemporaneous controls, and replacing experience
with age moves the canonical estimate's AR set across zero (Supplementary Table \ref*{tab:b11}).

CollegeDistance illustrates a protocol-driven change in diagnosis.
Shuffled folds give \(R^2_{\mathrm{oof},\pi}=0.78\); respecting the
\(43\)-group state proxy lowers it to \(-0.07\), revealing geographic leakage. Flexible-learner estimates
remain near zero with AR sets spanning both signs. Separate nuisance
selection removes the disagreement produced by coupled learners. With cross-fitting folds defined by the state proxy, all outcome learners have negative out-of-fold \(R^2\) (Supplementary \ref*{app:d4}).

Angrist--Krueger extends the inference comparison to multiple
instruments. Three quarter-of-birth instruments give a schooling
coefficient near \(0.108\), with bounded, similar AR, score, and
conditional-QLR sets (Supplementary Table \ref*{tab:b6}). With thirty quarter-of-birth-by-year
interactions, two-stage least squares moves toward OLS with apparent
Wald precision, while identification-robust sets are uninformative.
At \(51\) state-of-birth clusters, the thirty-instrument critical
value approaches the clustered AR tail statistic's cluster-count
ceiling (Supplementary \ref*{c.5-set-extraction}). Full application details are in Supplementary \ref*{app:d2}--\ref*{app:d6}.

\subsection{Cross-application synthesis}\label{cross-cell-synthesis}

The applications show why point-estimate stability and first-stage
fit cannot substitute for effect inference. A stable estimate may
accompany a wide or unbounded AR set, and a strength change must be
traced through folds and nuisance specification before being attributed
to instrument learning. Diagnostics identify plausible sources of
sensitivity; AR sets describe compatible effect values under the
maintained model and IV assumptions.

\begin{figure}[tbp]
\centering
\includegraphics[width=1.0\linewidth]{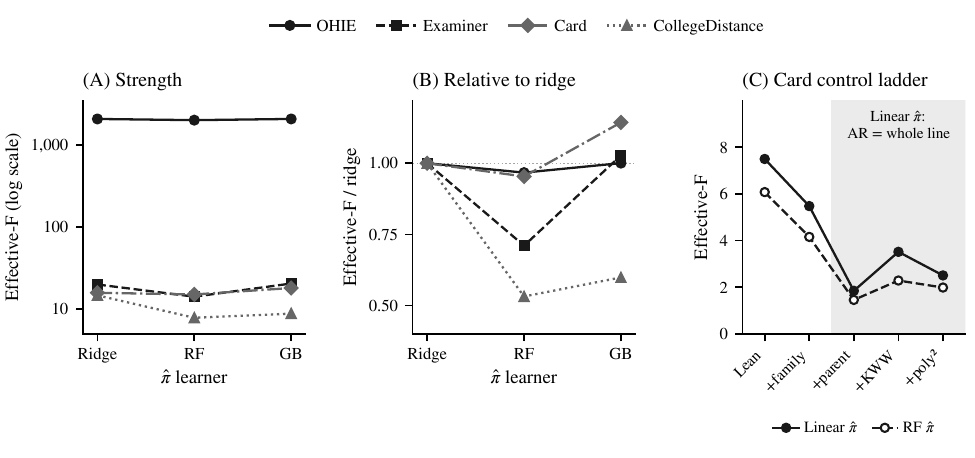}
\caption{Empirical comparisons of residualized strength.
(A) Effective-\(F\) across instrument-residualization learners;
(B) strength relative to each application's ridge baseline;
(C) the Card control ladder with sample and instrument held fixed.
Panels A--B compare learner sensitivity within each application;
levels across applications also reflect different designs.
Panel C illustrates changes in concentration, not a ranking of
causal specifications.}
\label{fig:fig1}
\end{figure}

Figure~\ref{fig:fig1} summarizes learner sensitivity and the fixed-sample
control ladder. Differences across populations and designs prevent
cross-application strength levels from isolating a common mechanism.
Section~\ref{nuisance-specification-and-residualized-strength} supplies the complementary comparison under known truth,
tracing the nuisance-specification reversal through the treatment residual.

\section{Discussion and Conclusion}\label{discussion}

The population result and the empirical diagnostics answer different
questions. Proposition~\ref{prop:concentration} ranks instrument residualizations for a fixed
conditioning set and a correctly specified treatment nuisance.
The applications assess why realized estimates and confidence sets
vary when learners, folds, or controls change. Their diagnostic
patterns support specific interpretations, but do not uniquely
identify a source of weakness or validate the IV assumptions.

The concentration result is related to classical optimal-instruments
theory \citep{amemiya1974,chamberlain1987,newey1990}, which chooses an
instrument transformation \(h(Z,W)\) to attain an efficiency bound
under strong identification. Here the instrument is fixed and the
choice is its residualization \(\pi\). Within this class, the two
criteria coincide under conditional homoskedasticity of the score,
where \(\operatorname{Var}(\hat\tau)\propto1/(nC)\) is exact
(Supplementary \ref*{a.1-proposition-1-sources-and-proof-sketch}). Under bounded conditional heteroskedasticity,
\(C\) ranks identification strength up to constants rather than
exactly tracking the sandwich variance.

The scope of inference is limited by both the target model and the
sampling assumptions. We maintain a homogeneous-coefficient PLIV
approximation; under heterogeneity, the causal interpretation depends
on the treatment and instrument, as Table~\ref{tab:4} records.
The clustered validity results are asymptotic in the number of
independent clusters and require nuisance-rate conditions. They do
not guarantee finite-sample coverage for sparse generated instruments
or many instruments with few clusters. Section~\ref{monte-carlo-studies} documents coverage
shortfalls as well as near-nominal performance. Moreover, coverage
at the true parameter does not make an unbounded set informative.

Two extensions remain useful. The cross-fitted AR moment contains a
second-order product of nuisance errors scaled by the tested effect,
so rate conditions can become more demanding at distant candidate
values. A locally robust corrected moment \citep{chernozhukov2022} could address these terms directly.
The identification-aware selector also warrants further study: its
validity conditions and its finite-sample accuracy are distinct,
and the present paired experiments do not establish an accuracy gain.

\label{conclusion}

Flexible instrument residualization has no inherent population
concentration cost under the maintained model and a correctly
specified treatment nuisance. In empirical work, this benchmark
supports joint reporting of nuisance diagnostics, residualized
strength, and identification-robust confidence sets, with the
estimand and IV assumptions stated explicitly.

\section*{Acknowledgements}
I am grateful to Di Zhang for carefully reading the manuscript and providing
detailed, constructive comments. I also thank Tamer {\c{C}}etin for sharing his
manuscripts and commenting on earlier drafts of this paper. All remaining errors
are my own.

\section*{Data and code availability}
The replication materials are not part of this posting. They contain the
analysis code, the frozen numerical results, scripts that rebuild the result
tables and statistical figures, the LaTeX tables and figure files, the software
requirements, and instructions for obtaining the input data. They will be
deposited in a public repository with a persistent identifier upon publication
of the article. Until then, they and the Supplementary Material cited in the
text are available from the author on request. The input data are not redistributed. The CollegeDistance, Card,
and Angrist--Krueger data are publicly available. The Oregon Health Insurance
Experiment public-use data \citep{finkelstein2013ohie} are available from the
National Bureau of Economic Research under its data use agreement. The examiner
data are available from openICPSR (project 209883, version V1,
\url{https://doi.org/10.3886/E209883V1}) under its terms of use, as part of the
deposit accompanying \citet{chyn2025}.

\section*{Declaration of generative AI and AI-assisted technologies in the manuscript preparation process}
During the preparation of this work, the author used OpenAI ChatGPT and Codex to
assist with manuscript organization, language revision, and LaTeX formatting.
The author reviewed and edited the assisted content and takes responsibility for
the final manuscript.

\bigskip
\setlength{\bibsep}{4pt}
\renewcommand{\bibsection}{\section*{References}\phantomsection\label{sec:references}}
\bibliographystyle{plainnat-doi}
\bibliography{references}

\end{document}